\documentclass[aip,reprint,showpacs,superscriptaddress]{revtex4-1}  

\usepackage[T1]{fontenc}
\usepackage[latin9]{inputenc}
\usepackage{units}
\usepackage{amsmath,amssymb,bm}
\usepackage{graphicx}
\usepackage{esint}
\usepackage[caption=false]{subfig}

\usepackage{bookman}
\usepackage{times}
\usepackage{ctable}
\usepackage{booktabs}
\usepackage{multirow}
\usepackage{floatrow}
\usepackage{wrapfig}
\usepackage{hyperref}

\usepackage{graphicx}
\usepackage{dcolumn}
\usepackage{bm}

\newcommand\T{\rule{0pt}{2.6ex}}       
\newcommand\B{\rule[-1.2ex]{0pt}{0pt}} 

\begin{document}


\title{Nonequilibrium phonon effects in midinfrared quantum cascade lasers}

\author{Y. B. Shi}\email{yshi9@wisc.edu}
\author{I. Knezevic}\email{knezevic@engr.wisc.edu}
\affiliation{Department of Electrical and Computer Engineering, University of Wisconsin - Madison, Madison, Wisconsin 53706-1691, USA}

\date{\today}

\begin{abstract}
We investigate the effects of nonequilibrium phonon dynamics on the operation of a GaAs-based midinfrared quantum cascade laser over a range of temperatures (77--300 K) via a coupled ensemble Monte Carlo simulation of electron and optical-phonon systems. Nonequilibrium phonon effects are shown to be important below 200 K. At low temperatures, nonequilibrium phonons enhance injection selectivity and efficiency by drastically increasing the rate of interstage electron scattering from the lowest injector state to the next-stage upper lasing level via optical-phonon absorption. As a result, the current density and modal gain at a given field are higher and the threshold current density lower and considerably closer to experiment than results obtained with thermal phonons. By amplifying phonon absorption, nonequilibrium phonons also hinder electron energy relaxation and lead to elevated electronic temperatures.
\end{abstract}

\maketitle


\section{Introduction}\label{sec:intro}

Quantum cascade lasers (QCLs) are electrically driven, unipolar semiconductor devices that achieve lasing by electronic transitions between discrete subbands formed due to confinement.\cite{Faist1994} Conventional midinfared (mid-IR) and resonant-phonon (RP) THz QCLs make use of the electron emission of optical phonons to assist the depopulation of the lower lasing level and help achieve population inversion. An electron typically generates one phonon per stage in RP THz QCLs, \cite{Vitiello2012,Iotti2013} and as many as 4-6 in mid-IR QCLs. \cite{Spagnolo2002} The electronic system in QCLs dissipates much of the energy received from the external field through the emission of optical phonons. \cite{Paulavicius1998} Phonon generation is typically faster than their anharmonic decay into acoustic modes, \cite{Paulavicius1998,Scamarcio2008,Iotti2010} so QCL laser operation is accompanied by a considerable population of nonequilibrium phonons. The presence of nonequilibrium optical phonons  has been experimentally observed using the combination of microprobe photoluminescence and Stokes/anti-Stokes Raman spectroscopy in the active region of both mid-IR \cite{Spagnolo2002} and RP THz QCLs. \cite{Vitiello2012,Scamarcio2013} Nonequilibrium phonons interact with the electronic system and can affect the device electronic and optical characteristics. The impact on mid-IR QCLs has been theoretically investigated using rate equations, \cite{Slivken1999} as well as the ensemble Monte Carlo (EMC) approach,\cite{Paulavicius1998,Compagnone2002} using a subset of subbands with assumed injection and extraction conditions. EMC has also been applied to analyze the influence of excess optical phonons on RP THz QCLs.\cite{Lu2006,Jirauschek2008,Iotti2010,Iotti2013}

In this paper, we study the effects of nonequilibrium phonon dynamics on the electron transport and laser performance in mid-IR QCLs over a range of temperatures (77--300 K) by means of an EMC solution to the coupled Boltzmann transport equations (BTEs) for electrons and longitudinal optical (LO) phonons. As an example, we consider the well-known 9-$\mu$m GaAs/AlGa$_{0.45}$As$_{0.55}$ mid-IR QCL \cite{Page2001} based on a conventional three-well active region design (Fig. \ref{Fig:BandDiagram}).  However, the technique is broadly applicable (a similar technique was recently used for RP THz QCLs\cite{Iotti2010,Iotti2013}) and we focus on the phenomena that are general for mid-IR QCLs. We show that nonequilibrium phonons are very important at low temperatures, for this particular design below 200 K, while their effects on QCL performance are negligible at higher temperatures. A key phenomenon is amplified interstage electron scattering with phonon absorption between the lowest injector and next-stage upper lasing levels, which leads to selectively enhanced injection and increased current and population inversion up to high fields, and also to threshold current densities lower and closer to experiment than the results of calculation with thermal phonons. Nonequilibrium phonons also result in enhanced electronic subband temperatures, as enhanced absorption of phonons effectively impedes electron energy relaxation.


This paper is organized as follows. In Sec. \ref{sec:theory}, we present the theoretical framework based on the coupled Boltzmann equations for electrons and LO phonons. In Sec. \ref{sec:results}, we show the calculated laser characteristics with and without nonequilibrium phonons and elucidate their microscopic underpinnings. We conclude with Sec. \ref{sec:conclusion}.

\section{Theoretical Framework}\label{sec:theory}

\begin{figure}
\centering\includegraphics[height=2 in]{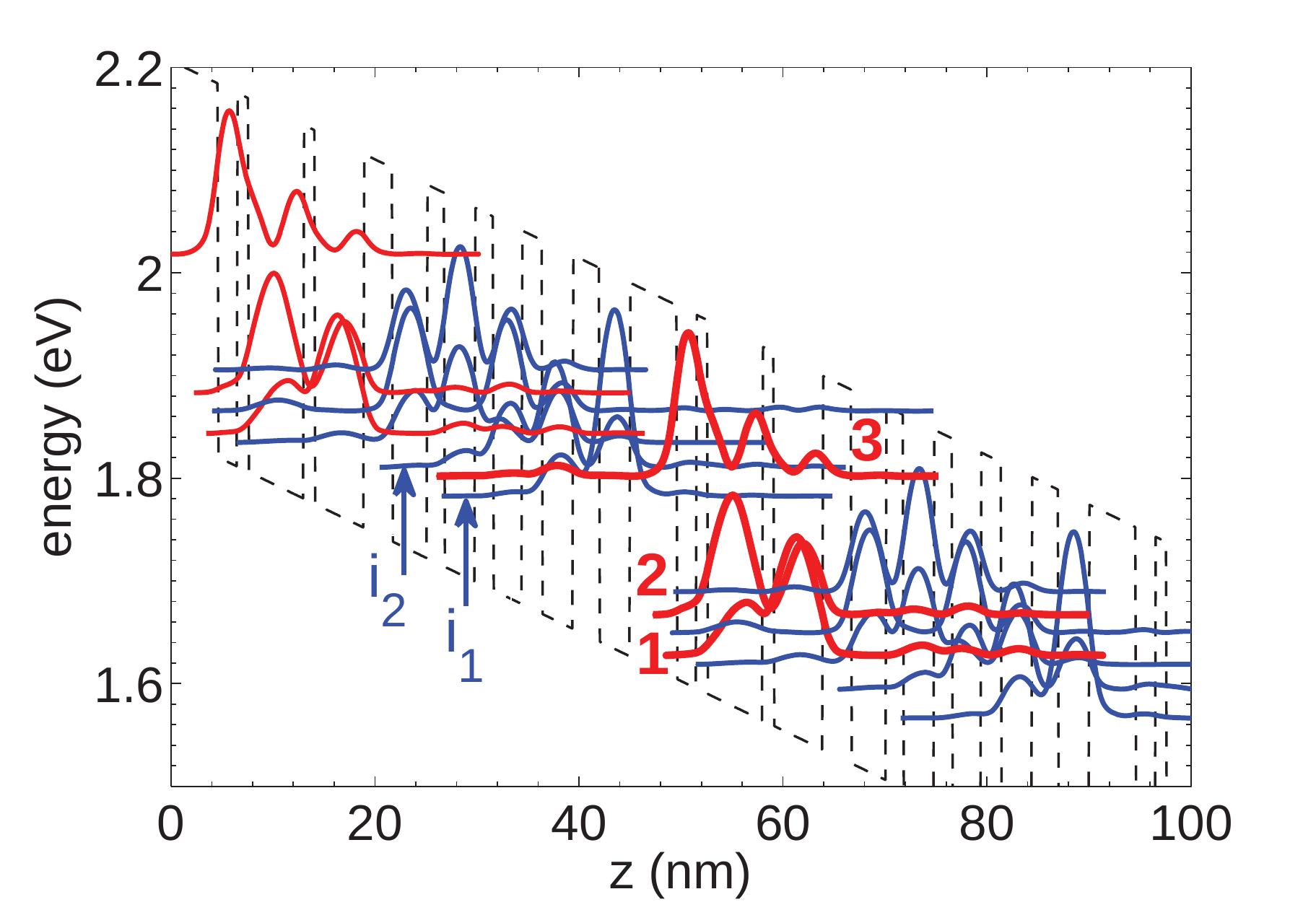}
\caption{\label{Fig:BandDiagram}
Energy levels and wavefunction moduli squared of $\Gamma$-valley subbands in two adjacent stages of the simulated GaAs/AlGaAs-based structure. The bold red curves denote the active region states (1, 2, and 3 represent the ground state and the lower and upper lasing levels, respectively). The blue curves represent injector states, with $i_1$ and $i_2$ denoting the lowest two. }
\end{figure}

The simulated structure is the well-characterized 9-$\mu$m GaAs/AlGa$_{0.45}$As$_{0.55}$ mid-IR QCL from Ref. [\onlinecite{Page2001}]. The layer thickness (in \AA) in one stage, starting from the injection barrier, is \textbf{46}/19/\textbf{11}/54/\textbf{11}/48/\textbf{28}/34/\textbf{17}/30/\underline{\textbf{18}}/\underline{28}/\underline{\textbf{20}}/\underline{30}/\textbf{26}/30.
The bold script denotes barriers, the normal script are wells, and the underscored regions are \emph{n}-type doped with a sheet density of $N_s$ = 3.8$\times$10$^{11}$ cm$^{-2}$.
Figure~\ref{Fig:BandDiagram} shows the subband energy levels and wavefunction moduli squared in two stages of the simulated structure at a field of $48\,\mathrm{kV/cm}$. The red curves represent the active region states, while the blue curves represent the injector states. The radiative transition happens between the upper lasing level 3 and lower lasing level 2, while state 1 is the ground state designed to depopulate level 2. Only the $\Gamma$-valley states are included in the simulation, as it has been demonstrated that intervalley leakage in this particular QCL design is negligible. \cite{Gao2006,Gao2007}

Within the semiclassical framework, the dynamics of the interacting electron and LO phonon systems  in QCLs can be described by the coupled BTEs:\cite{Iotti2010}
\begin{subequations}\label{eq:BTEs}
\begin{eqnarray}
\frac{\mathrm{d} f_{\mathrm{i},\textbf{k}_\parallel}}{\mathrm{d} t} &=&
\left.\frac{\mathrm{d} f_{\mathrm{i},\textbf{k}_\parallel}}{\mathrm{d} t}\right|_{\mathrm{e-ph}}
+ \left.\frac{\mathrm{d} f_{\mathrm{i},\textbf{k}_\parallel}}{\mathrm{d} t}\right|_{\mathrm{e-e}},\label{eq:electronBTE} \\
\frac{\mathrm{d}N_{\textbf{q}}}{\mathrm{d} t} &=&\left.\frac{\mathrm{d} N_{\textbf{q}}}{\mathrm{d} t}\right|_{\mathrm{ph-e}}
-\frac{N_{\mathbf{q}} - N_0}{\tau}.\label{eq:phononBTE}
\end{eqnarray}
\end{subequations}
Here, $f_{\mathrm{i},\textbf{k}_\parallel}$ is the electron distribution function for subband $\mathrm{i}$ and in-plane wave vector $\textbf{k}_\parallel$. $N_\textbf{q}$ is the optical phonon occupation number for wave vector $\textbf{q}$ and is also often referred to as the phonon distribution function; we will use the two terms interchangeably. The coupled BTEs (\ref{eq:BTEs}) are solved by the particle-based EMC approach.\cite{Gao2007,Jirauschek2008,Iotti2010,Iotti2013,JirauschekKubisAPR2014} The equations are coupled through the electron-phonon collision integrals $\left(\mathrm{d} f_{\mathrm{i},\textbf{k}_\parallel}/\mathrm{d} t)\right|_{\mathrm{e-ph}}$ and $\left(\mathrm{d} N_{\textbf{q}}/\mathrm{d} t)\right|_{\mathrm{ph-e}}$, which can be evaluated using Fermi's Golden Rule.\cite{Lundstrom92}
In the electronic transport EMC, we account for intrasuband and intersubband electron--LO phonon scattering, as well as for electron--electron scattering within the static random-phase approximation. \cite{Goodnick1988} We assume bulk phonons, since phonon confinement was shown to negligibly affect the electron--LO phonon scattering rates in mid-IR QCLs,\cite{Gao2008} and adopt  periodic boundary conditions.\cite{Gao2007,Iotti2010}

The phonon BTE (\ref{eq:phononBTE}) has been simplified by applying the relaxation time approximation to describe the anharmonic decay of a LO phonon into two longitudial acoustic (LA) phonons; the corresponding relaxation time $\tau$ is analytically derived \cite{Usher1994} and is of order picoseconds (e.g., $\tau=8.5\,\mathrm{ps}$ at 77 K).  $N_0=\left[\exp(\hbar\omega_0/k_B T_L)-1\right]^{-1}$ is the thermal equilibrium phonon occupation number for assumed dispersionless optical phonons of energy $\hbar\omega_0=36\,\mathrm{meV}$ at lattice temperature $T_L$ ($k_B$ is the Boltzmann constant). We capture the time evolution of the nonequilibrium phonon distribution $N_{\textbf{q}}$ (\ref{eq:phononBTE}) via a histogram \cite{Lugli1989} generated over a discrete set of $q_\parallel$ (magnitude of the in-plane $\textbf{q}_{\parallel}$) and $q_z$ (the cross-plane phonon momentum), where a `numerical' phonon is added/deleted every time an LO  phonon is emitted/absorbed in the electronic EMC. At the end of each time step, the remaining nonequilibrium LO phonons are allowed to randomly decay based on the anharmonic decay time $\tau$. $N_{\textbf{q}}$ is obtained from the histogram by taking into account the phonon density of states, and is used to calculate the updated electron-phonon scattering rates, which are propotional to $N_{\textbf{q}}$ for phonon absorption and to $N_{\textbf{q}}+1$ for emission, for use during the next time step.

\begin{figure}
\centering\includegraphics[height=2 in]{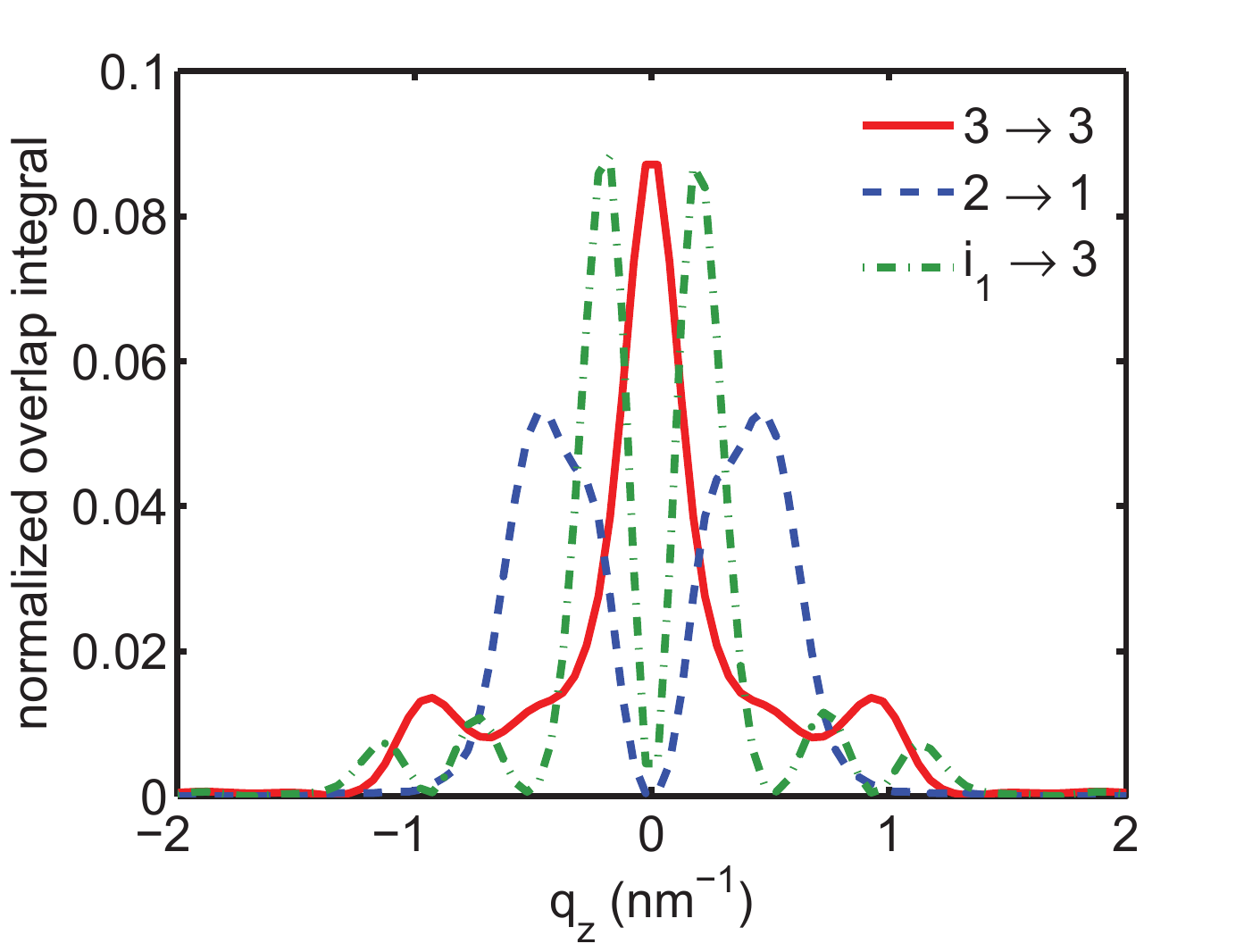}
\caption{\label{Fig:OverlapInt}
Normalized overlap integral $|\mathcal{I}_{\mathrm{if}}|^2$ from Eq. (\ref{eq:OverlapIntegral}) versus cross-plane phonon wave vector $q_z$ for several transitions (intersubband $i_1\rightarrow 3$ and $2\rightarrow 1$;  intrasubband $3\rightarrow 3$).}
\end{figure}

While the in-plane wave vector $\textbf{q}_\parallel$ of a bulk optical phonon involved in an electron-phonon scattering event is determined from in-plane momentum conservation, the cross-plane component $q_z$ is not strictly conserved because of electron confinement. \cite{Price1981,Ridley1982,Riddoch1983} Previously, the momentum conservation approximation (MCA)\cite{Ridley1982,Lu2006} and the uniform phonon distribution approximation with broadening \cite{Lugli1989} have been applied to this problem (the latter on a quantum well, not a QCL); the MCA underestimates the phonon mode exchange,\cite{Paulavicius1998} while the broadening length is difficult to determine in QCLs. Here, we note that the probability distribution of $q_z$ of phonons generated in transitions from initial subband $\mathrm{i}$ to final subband $\mathrm{f}$ is proportional to the the overlap integral (OI) between the initial and final states [$\psi_\mathrm{i}(z)$ and $\psi_\mathrm{f}(z)$, respectively] over the simulation domain of length $L$ (the OI is also involved in the electron-phonon scattering rate calculation):
\begin{equation}\label{eq:OverlapIntegral}
|\mathcal{I}_{\mathrm{if}}(q_{z})|^2 = \int_0^L \mathrm{d}z
\psi_f^*(z) \psi_i(z) e^{-i q_z z}.
\end{equation}
In this work, we use a random variable whose probability distribution follows the \textit{normalized OI}, depicted in Fig. \ref{Fig:OverlapInt}, to determine the $q_z$ of a phonon involved in the $\mathrm{i} \rightarrow \mathrm{f}$ electron transition. The peak width of the normalized OI indicates how sharply the cross-plane momentum is conserved,  \cite{Riddoch1983} while the peak positions depend on the subband energy separation. Note how the intrasubband OIs are strongly peaked at $q_z = 0$, while the intersubband OIs are zero at $q_z = 0$ (Fig. \ref{Fig:OverlapInt}). The area of overlap between the regions under the OI curves corresponding to different transitions indicates the mutual optical-phonon mode exchange,\cite{Paulavicius1998} i.e. quantifies how frequently an optical phonon created during one transition can participate in a different one.

\section{Results and Discussion}\label{sec:results}


\subsection{Laser characteristics with and without nonequilibrium phonons}\label{sec:laser characteristics}

\begin{figure}
\centering\includegraphics[height=2 in]{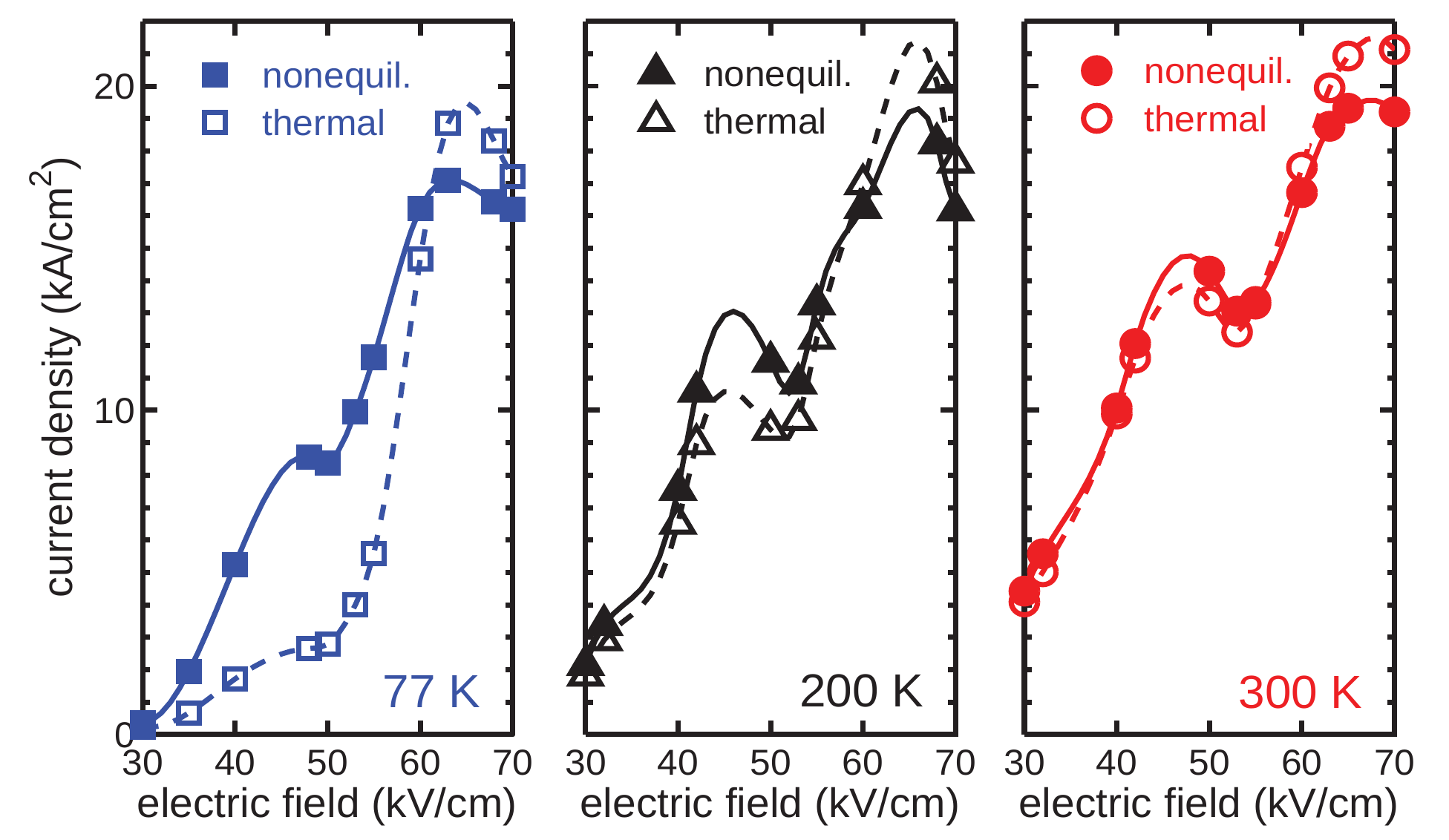}
\caption{Current density versus applied electric field obtained from the simulations with nonequilibrium (solid curves) and thermal (dashed curves) phonons at 77, 200, and 300 K.}\label{Fig:IVcurves}
\end{figure}

\begin{figure}
\centering\includegraphics[height=2 in]{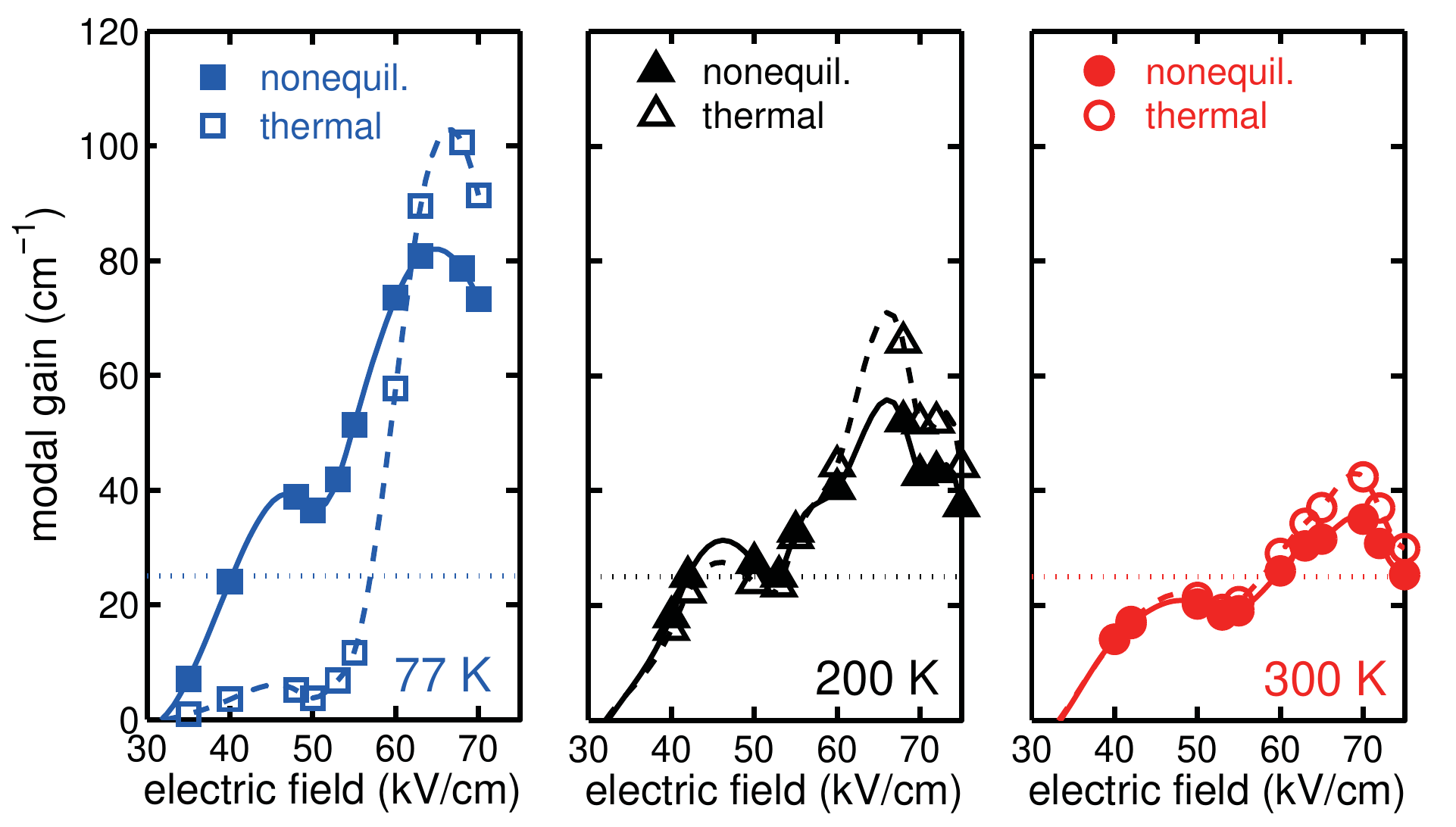}
\caption{Modal gain as a function of electric field, obtained from the simulations with nonequilibrium (solid curves) and thermal (dashed curves) phonons at 77, 200, and 300 K. The horizontal dashed line denotes the estimated total loss of $\alpha_{\mathrm{tot}}=25\,\mathrm{cm}^{-1}$.}\label{Fig:modal gain vs E field}
\end{figure}

\begin{figure}
\centering\includegraphics[height=2 in]{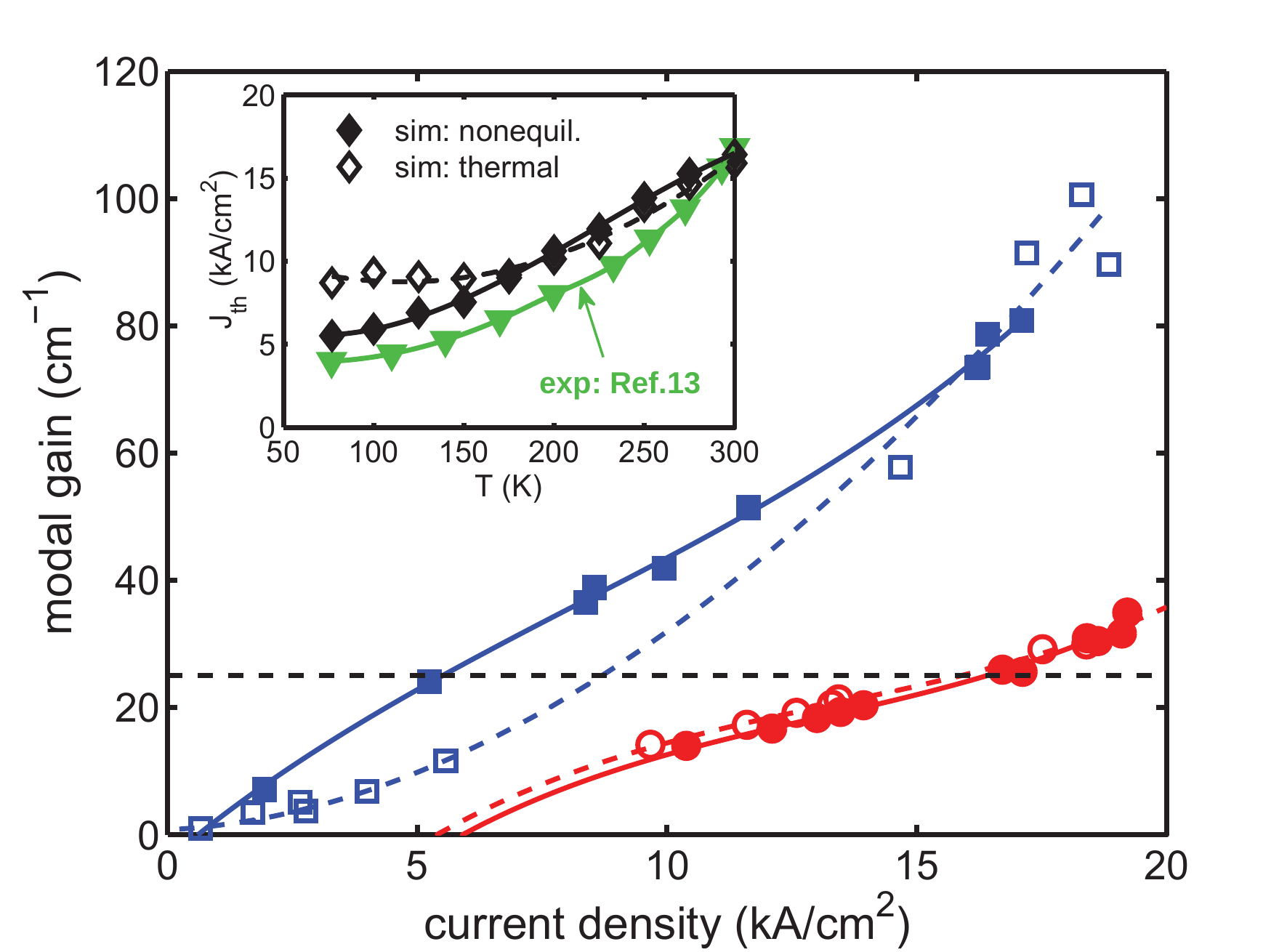}
\caption{Modal gain as a function of current density, obtained from the simulations with nonequilibrium (solid curves) and thermal (dashed curves) phonons at 77 and 300 K.The horizontal dashed line denotes the estimated total loss of $\alpha_{\mathrm{tot}}=25\,\mathrm{cm}^{-1}$. Inset: Threshold current density vs lattice temperature, as calculated with nonequilibrium (black solid curve) and thermal (black dashed curve) phonons, and as obtained from experiment [\onlinecite{Page2001}] (green curve). }\label{Fig:modal gain vs current}
\end{figure}

In Figs. \ref{Fig:IVcurves}, \ref{Fig:modal gain vs E field}, and \ref{Fig:modal gain vs current}  we present several laser characteristics, important in experiment, calculated with and without nonequilibrium phonons and at temperatures of 77, 200, and 300 K: the $J\textrm{--}F$ curves (Fig. \ref{Fig:IVcurves}), modal gain versus electric field (Fig. \ref{Fig:modal gain vs E field}), and modal gain versus current density (Fig. \ref{Fig:modal gain vs current}). Modal gain is calculated as \cite{Mircetic2005} $G_m=\frac{4\pi e^2{\langle z_{32}\rangle}^2\Gamma_w \Delta n}{2 \epsilon_0 \underline{n}\gamma_{32}L_p \lambda}$, with the following values:\cite{Page2001,SirtoriAPL1999,GaoJAP2007b} the dipole matrix element $\langle z_{32}\rangle=1.7\,\mathrm{nm}$, optical mode refractive index $\underline{n}=3.21$, full width at half maximum $ \gamma_{32}(T_L)\approx 8.68\,\mathrm{meV}+0.045\,\mathrm{meV/K}\times T_L$, stage length $L_p=45\,\mathrm{nm}$, wavelength $\lambda = 9\,\mathrm{\mu m}$, waveguide confinement factor $\Gamma_w = 0.31$. Population inversion $\Delta n$ is obtained from the EMC simulation. Waveguide loss $\alpha_w=20\,\mathrm{cm}^{-1}$ and  mirror loss $\alpha_m=5\,\mathrm{cm}^{-1}$ are assumed,\cite{SirtoriAPL1999,GaoJAP2007b}  yielding a total loss estimate of $\alpha_{\mathrm{tot}}=25\,\mathrm{cm}^{-1}$.

The calculated laser characteristics reveal several important features, whose microscopic underpinnings we discuss below. In Figs. \ref{Fig:modal gain vs E field} and \ref{Fig:modal gain vs current} we see that the current density and gain at a given field are considerably higher with nonequilibrium phonons than thermal ones at 77 K, while the difference is small at 200 K and barely perceptible at 300 K; this trend holds up to high fields ($>60\,\mathrm{kV/cm}$). Furthermore, in the inset of Fig. \ref{Fig:modal gain vs current}, we see that, below 200 K, the inclusion of nonequilibrium phonons in the calculation gives lower threshold current densities, considerably closer to experiment [\onlinecite{Page2001}] with very similar design parameters than the results with thermal phonons. As we show below, these features have the same underlying microscopic mechanism.


\subsection{How nonequilibrium phonons affect QCL characteristics -- the microscopic picture}\label{sec:microscopic mechanism}

\begin{figure}
\centering\includegraphics[height=2 in]{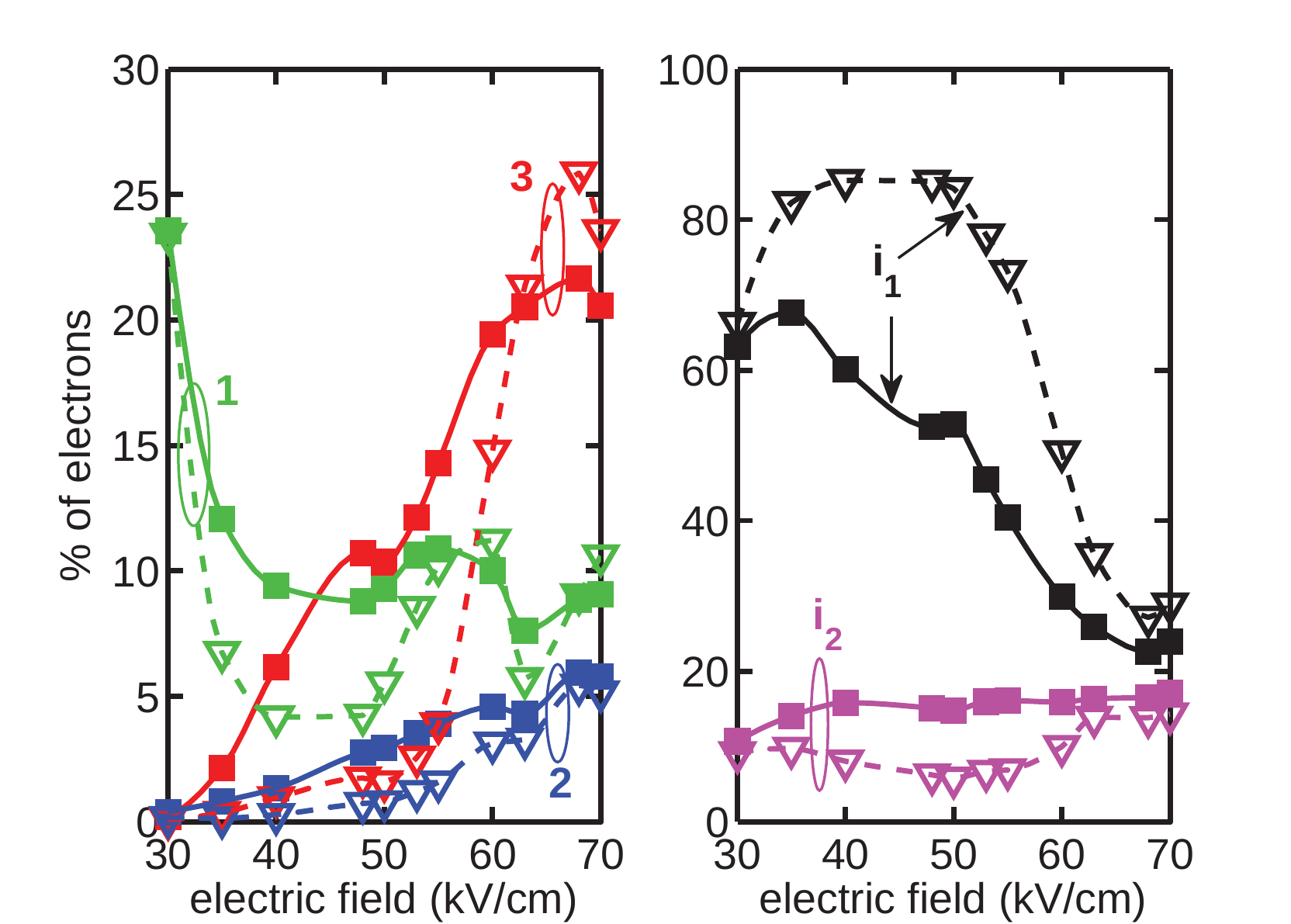}
\caption{Population of the active region levels 3, and 2, and 1 (left panel) and the bottom two injector levels $i_2$ and $i_1$ (right panel) versus applied electric field obtained with nonequilibrium (solid curves) and thermal (dashed curves) phonons at 77 K. }\label{Fig:subband population 77 K}
\end{figure}

The enhanced modal gain in Fig. \ref{Fig:modal gain vs current} stems from enhanced population inversion in the presence of nonequilibrium phonons. In Fig. \ref{Fig:subband population 77 K}, we present the percentage population of active region levels 3, 2, and 1 (left panel) and the bottom two injector states $i_2$ and $i_1$ (right panel) as a function of electric field at 77 K, with and without nonequilibrium phonons (solid and dashed curves, respectively). We see that the population of the upper lasing level, $n_3$, is considerably higher with nonequilibrium phonons than it is equilibrium phonons, while the population of the lower lasing level, $n_2$, is only slightly enhanced by them; as a result, the population inversion $\Delta n=n_3-n_2$ and modal gain are considerably higher with than without nonequilibrium phonons.

\begin{figure}\centering\includegraphics[width=\columnwidth]{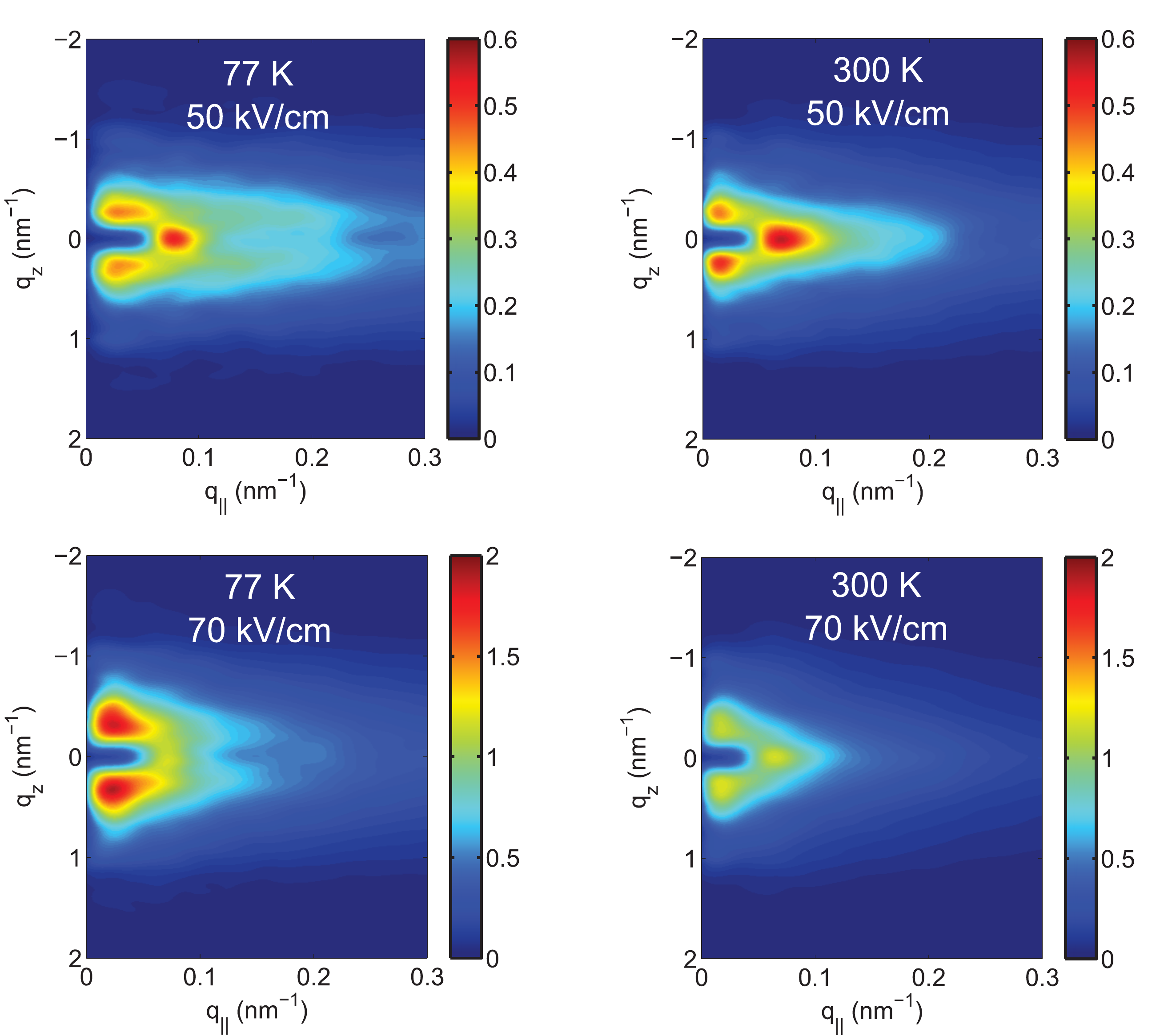}
\caption{Nonequilibrium phonon occupation number, $N_{\textbf{q}}-N_0$, presented via color (red -- high, blue -- low) at temperatures of 77 K and 300 K and fields of 50 kV/cm and 70 kV/cm. Note the different color bars that correspond to different fields. }\label{Fig:noneq phonons}
\end{figure}

How do nonequilibrium phonons enable this increase in the population inversion? At low temperatures, the occupation number of thermal phonons is very small [$N_0 (77\,\mathrm{K})\approx 0.06$], whereas the occupation number of nonequilibrium phonons can be one to two orders of magnitude higher, depending on the field. Figure \ref{Fig:noneq phonons} shows the nonequilibrium phonon occupation, $N_{\textbf{q}}-N_0$, at different temperatures (77 and 300 K) and fields (50 kV/cm and 70 kV/cm); note that the color bars differ at the two fields. Nonequilibrium phonon occupation number is appreciable in a small segment of the Brillouin zone. Nonequilibrium phonons elevate the rates of both absorption (proportional to $N_{\textbf{q}}$) and emission (proportional to $N_{\textbf{q}}+1$) of phonons by electrons, but the effect is particularly dramatic on absorption. In Table I, we show the average electron lifetimes (in ps) for different transitions among the active region and lowest two injector states at 77 K and 50 kV/cm, with and without nonequilibrium phonons. The intersubband rates most highly enhanced (i.e. the lifetimes most highly reduced) by nonequilibrium phonons, by a factor of roughly 40, are $i_1\rightleftarrows 3$ and $i_2\rightleftarrows 3$ (see Fig. \ref{Fig:BandDiagram}). When we consider the high population of $i_1$, and the low population of $i_2$ and $3$ (Fig. \ref{Fig:subband population 77 K}), it becomes clear that the current component most enhanced by the presence of nonequilibrium phonons corresponds to the $i_1\rightarrow 3$ transition with phonon absorption (this current is proportional to $n_{i_1}/\tau_{i_1\rightarrow 3}$, $n_{i_1}$ being the population of $i_1$). At the same time, the impact of nonequilibrium phonons on the parasitic injection channels is relatively small because of the high energy separation (see Table I). Therefore, nonequilibrium phonons improve the injection selectivity by preferentially amplifying the rate of interstage injector--upper lasing level electron scattering with phonon absorption, which results in higher gain and a lower threshold current density. The enhancement in current that is visible in Fig. \ref{Fig:IVcurves} up to 60 kV/cm has the same underlying reason:  $i_1\rightarrow 3$ interstage scattering with phonon absorption, amplified in the presence of nonequilibrium phonons.

Enhancement in the electron absorption of phonons, enabled by nonequilibrium phonons, has another manifestation. At all temperatures, the current density at high fields (> 60 kV/cm) is lower with nonequilibrium than with thermal phonons. The reason is that, with increasing field, $i_1$ moves upward with respect to $3$, crossing it at about 60 kV/cm. Therefore, at high fields, current due to backscattering $3\rightarrow i_1$ with phonon absorption is the component most amplified by nonequilibrium phonons, and since it is negative, we see an overall lower current with nonequilibrium phonons.

\begin{table}%
\caption{Average relaxation time (in ps) at 77 K and 50 kV/cm among injector and active region states
($i_2$, $i_1$, 3, 2, and 1; see Fig. \ref{Fig:BandDiagram}). Rows correspond to initial subband, columns to final. Normal script corresponds to thermal phonons, boldface to nonequilibrium phonons.}
\label{Table2}\centering%
\begin{ruledtabular}
\begin{tabular}{c@{\hspace{2em}}cc@{\hspace{1em}}cc@{\hspace{1em}}cc@{\hspace{1em}}cc@{\hspace{1em}}cc}
{}&\multicolumn{2}{c}{$i_2$}&\multicolumn{2}{c}{$i_1$}&\multicolumn{2}{c}{$3$}
&\multicolumn{2}{c}{$2$}&\multicolumn{2}{c}{$1$}\T\B\\
\midrule
$i_2$ & 27 & \bf{0.7} & 64 & \bf{1.7} & 478 & \bf{13} & 68 & \bf{65} & 103& \bf{101}\\
\midrule

$i_1$ & 69	& \bf{2.0}	& 19	&   \bf{0.5}	& 349	& \bf{8.6}	& 62	& \bf{59}	& 108 & \bf{104}\\
\midrule

$3$ & 347 & \bf{8.8}	& 495	&\bf{13} & 26 & 	\bf{0.7} & 	1.7	 & \bf{1.6}	 & 2.7 & 	 \bf{2.6}\\
\midrule

$2$ & 704 & 	\bf{144}	 & 3453	 & \bf{167}	 & 47	 & \bf{4.1}	 & 13	 & \bf{0.5}	 & 0.4	 & \bf{0.3}\\
\midrule

$1$ & 106 & 	\bf{79} &  597	 & \bf{104} & 	7.1 & 	\bf{2.8}	 & 30 & 	\bf{1.1} & 	23 & 	 \bf{0.6}
\end{tabular}
\end{ruledtabular}
\end{table}


\subsection{Electronic subband temperatures with and without nonequilibrium phonons}\label{sec:electronic temperature}

\begin{figure*}
\centering\includegraphics[width=\columnwidth]{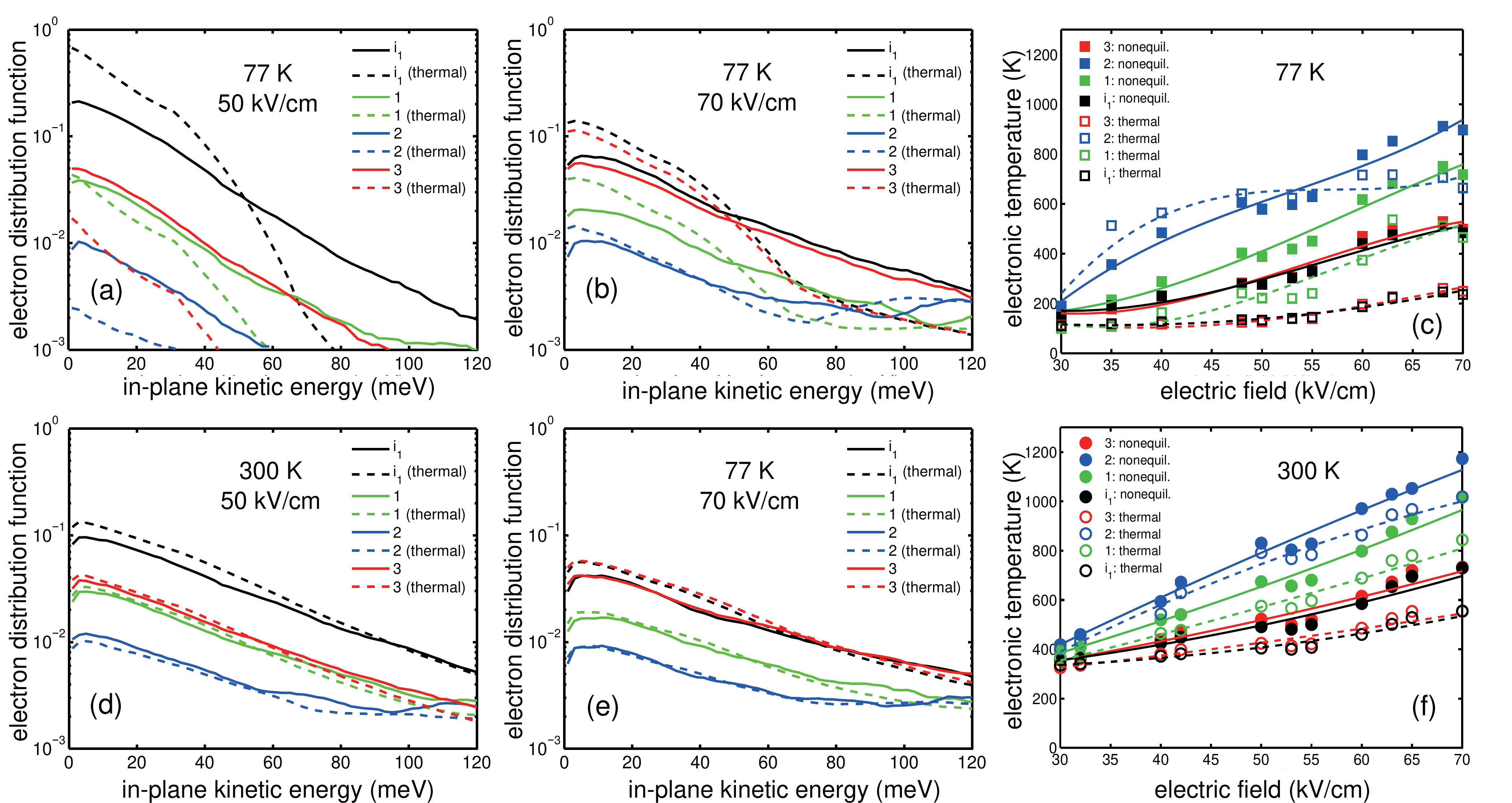}
\caption{(a,b,d,e) Population of the active region levels (3, 2, and 1) and the lowest injector state $i_1$ as a function of electron in-plane kinetic energy at the lattice temperature of 77 K and 300 K and at fields 50 kV/cm and 70 kV/cm, obtained with nonequilibrium (solid curves) and thermal (dashed curves) phonons. (c) \& (f) Electron temperature vs applied electric field at the lattice temperature of 77K and 300 K.  }\label{Fig:electron subbands}
\end{figure*}

In Figs. \ref{Fig:electron subbands}(a,b,d,e), we show the electronic distribution functions in different subbands at different temperatures (77 and 300 K) and fields (50 kV/cm and 70 kV/cm) with (solid curves) and without (dashed curves) nonequilibrium phonons. At 300 K [Figs. \ref{Fig:electron subbands}(d),\ref{Fig:electron subbands}(e)], with or without nonequilibrium phonons, the electrons are well thermalized; the distributions show a typical heated Maxwellian profile, a signature of strong bi-intrasubband electron-electron scattering, \cite{Iotti2001} as evidenced by the linear dependencies on the semilog plots in Fig. \ref{Fig:electron subbands}; the slope is $-1/k_BT_e$, where $T_e$ is the subband electron temperature. At 77 K [Figs. \ref{Fig:electron subbands}(a),\ref{Fig:electron subbands}(b)], nonequilibrium phonons aid in the thermalization, as the distributions are much closer to Maxwellian with than without nonequilibrium phonons; the long-energy tails present in the distributions are a signature of the high rate of phonon absorption. Indeed, amplified electron absorption of phonons impedes the energy relaxation of the electron system and results in higher electronic-subband temperatures $T_e$ with nonequilibrium than thermal phonons, as shown in the plots of $T_e$, calculated from average kinetic energy, vs. field at 77 K and 300 K [Figs. \ref {Fig:electron subbands}(c),\ref{Fig:electron subbands}(f)].

\section{Conclusion}\label{sec:conclusion}

We investigated the impact of nonequilibrium phonons on electron transport in a mid-IR GaAs-based QCL over a range of temperatures (77--300 K) using a coupled electron and phonon EMC technique that explicitly takes into account the phonon momentum distribution. The overarching message is that nonequilibrium phonons are important at temperatures below about 200 K and negligible otherwise. At low temperatures and in the presence of nonequilibrium phonons, the electron-LO phonon absorption rate increases by one to two orders of magnitude, and this microscopic phenomenon has several manifestations. Nonequilibrium phonons lead to a selective enhancement of injection from the lowest injector state into the upper lasing level via LO phonon absorption, which results in higher modal gain and current at a given field and a threshold current density lower and considerably closer to experiment than the calculation with equilibrium phonons. By amplifying phonon absorption, nonequilibrium phonons impede electron energy relaxation and lead to broader electron distributions and higher electronic temperatures than the simulation with thermal phonons.

\section*{Acknowledgment}
The authors thank D. Botez for comments on the manuscript. This work was funded by the U.S. Department of Energy, Office of Basic Energy Sciences, Division of Materials Sciences and Engineering under Award DE-SC0008712.

%

\end{document}